# BOOTSTRAP PETTITT TEST FOR DETECTING CHANGE POINT IN HYDROCLIMATOLOGICAL DATA: A CASE STUDY FOR ITAIPU HYDROELECTRIC PLANT IN BRAZIL

Luiza Chiarelli Conte[*]
Débora Missio Bayer[**]
Fábio Mariano Bayer[***]

## ABSTRACT

The Pettitt test has been widely used in climate change and hydrological analyzes. However, studies evidence difficulties of this test in detecting change points, especially in small samples. This study presents a bootstrap application of the Pettitt test, which is numerically compared with the classical Pettitt test by an extensive Monte Carlo simulation study. The proposed test outperforms the classical test in all simulated scenarios. An application of the tests is conducted in the historical series of naturalized flows of the Itaipu Hydroelectric plant in Brazil, where several studies have shown a change point in the 70s. When the series is split into shorter series, to simulate small sample actual situations, the proposed test is more powerful than the classical Pettitt test to detect the change point. The proposed test can be an important tool to detect abrupt changes in water availability, supporting hydroclimatological resources decision making.

## 1 INTRODUCTION

In environmental studies it is important to understand the behavior of some variables, with special interest in changes of climatological and hydrological data that could affect water availability or climate (Alley et al., 2003; Dailidiené et al., 2010; Fu et al., 2004; IPCC, 2014, 2007, 2001, 1995, 1992; Kundzewicz and Robson, 2000, 2004; Loucks et al., 2005; National Research Council, 2010, 2002; National Academies of Sciences and Medicine, 2016; Reis and Yilmaz, 2008; Rosenzweig et al., 2008; UNDP, 2007; Wang and Qin, 2017; WMO, 1988, 2016; Zhang et al., 2007). Identifying changes and their causes can be very important in an infrastructure project, water management,

[*] Programa de Pós-Graduação em Engenharia Civil, Universidade Federal de Santa Maria, Santa Maria, Brazil.

[**] Departamento de Engenharia Sanitária e Ambiental, Universidade Federal de Santa Maria, Santa Maria, Brazil.

[***] Departamento de Estatística and LACESM, Universidade Federal de Santa Maria, Santa Maria, Brazil.

Centro de Tecnologia, Av. Roraima, 1000, Cidade Universitária, Camobi, Santa Maria, Rio Grande do Sul, Brazil. E-mail: debora.bayer@ufsm.br

risk mitigation policies, and other hydroclimatological situation (Serinaldi and Kilsby, 2016). These changes can occur gradually (trend), abruptly (change point) or in more complex forms (Kundzewicz and Robson, 2000). As earlier the change in a hydroclimatic data set is identified, greater are the benefits for climate monitoring (Beaulieu et al., 2012). This way, it is important to investigate and improve change detection methods.

Studies using different change point detection methods have been developed to identify artificial or natural discontinuities and climatic changes in hydroclimatic data series (Beaulieu et al., 2012; Chandler and Scott, 2011; Kundzewicz and Robson, 2004). This change point can be defined as a point in time where any parameter of the data distribution, such as mean, median, variance, and/or autocorrelation, undergoes an abrupt change (Chandler and Scott, 2011; Kundzewicz and Robson, 2004). In abrupt change, the variable does not maintain a continuous average over long periods, and in the case of hydrological variables, this can be due to natural and/or anthropogenic changes (National Research Council, 2002). In Brazil, as in other countries, the land use and land cover have been changing, and native habitats have been converting to agro-pastoril land cover (Brannstrom et al., 2008; Nóbrega et al, 2018). These practices can affect the hydrological processes (Bosch and Hewlett, 1982; Andreàssian, 2004; Bayer and Collischonn, 2013; Brown et al 2013; Zang et al, 2017). In this sense, it is important to detect these changes in hydrological time series in order to associate them with their causes.

Several statistical tests have been developed to identify changes in a time series. Kundzewicz and Robson (2004) mention, among others, the Pettitt test (Pettitt, 1979), the Mann-Whitney test (Mann and Whitney, 1947), the CUSUM test (Buishand, 1982), the Kruskal-Wallis test (Kruskal and Wallis, 1952), the Student's t-test (Chandler and Scott, 2011), and the likelihood ratio test (Worsley, 1979). Among these tests, the Pettitt test has been widely used in hydroclimatological studies to identify change points in variables, such as rainfall or precipitation (Busuioc and von Storch, 1996; Karabörk et al., 2007; Liu et al., 2010; Rougé et al., 2013; Tarhule and Woo, 1998; Wijngaard et al., 2003), discharge or runoff (Jiang et al., 2011; Liu et al., 2010, 2012; Villarini et al., 2009; Villarini and Smith, 2010; Wang et al., 2013), base flow (Gao et al., 2015), temperature (Liu et al., 2012; Rougé et al., 2013; Suhaila and Yusop, 2017;Wijngaard et al., 2003), relative humidity (Liu et al., 2012), potential evapotranspiration (Zuo et al., 2012), and sunshine duration (Liu et al., 2012). In most of these studies, the aim was identifying the abrupt change to assess its causes. The Pettitt test becomes more advantageous because it is able to identify the point at which the change occurred in the data series, without the need of prior identification, besides being a nonparametric test (Mallakpour and Villarini, 2016) and quite flexible to hydroclimatic data.

Recent studies have been developed to evaluate the sensitivity of the Pettitt test in detecting changes in the mean, with performance analysis using Monte Carlo simulation (Mallakpour and Villarini, 2016; Xie et al., 2014). These studies have shown the Pettitt test performs well in large sample sizes, with a large magnitude of the change point and when the change point is near to the midpoint of the analyzed series. However, when these characteristics are not present in the series, especially in small sample sizes, the performance of the Pettitt test can be poor. In this sense, this paper proposes an improvement of the Pettitt test by means of the bootstrap method (Davison and Hinkley, 1997; Efron and Tibshirani, 1994). To evaluate the performance of the application of bootstrap method on Pettitt test (proposed test), we consider a Monte Carlo simulation study and an application to a real data set.

This article is organized as follows. Firstly, in Section 2, we describe the Pettitt test and the types of errors present in hypothesis testing. In Section 3, the proposed bootstrap application to the Pettitt test is presented. Section 4 presents the numerical results obtained from the comparison between the proposed test and the classical Pettitt test, via an extensive Monte Carlo simulation study. In Section 5 we present the classical and proposed tests application in historical series of naturalized flows of the Itaipu Hydroelectric, Brazil. Finally, Section 6 presents the conclusions and final remarks of the study.

## 2 PETTITT TEST

The Pettitt (Pettitt, 1979) method is a rank-based nonparametric statistical test used to detect change points present in a data series. Consider a sequence of random variables X1, X2, … , XT with change point at time $\tau$, which is divided in two sets with distinct distributions functions, such that F1(Xt) to t = 1, 2,…, $\tau$, and F2(Xt) to t = $\tau$+1,…, T.

Thus, the hypothesis to be tested are:

$$\begin{cases} H_0: F_1(X) = F_2(X) \ \text{(no change point)}, \\ H_1: F_1(X) \neq F_2(X) \ \text{(change point)}. \end{cases}$$

To detect the change point, the test uses the statistic Ut,T , similar to the Mann-Whitney test statistic (Mann and Whitney, 1947) for two samples, given by (Pettitt, 1979):

$$U_{t,T} = \sum_{i=1}^{t} \sum_{j=t+1}^{T} \text{sgn}(X_i - X_j), \qquad 1 \leq t < T,$$

Where

$$\text{sgn}(x) = \begin{cases} 1, & \text{if } x > 0, \\ 0, & \text{if } x = 0, \\ -1 & \text{if } x < 0. \end{cases}$$

The most probable change point $\tau$ will be the one that satisfies the following equation:

$$K_\tau = U_{\tau,T} = \max|U_{t,T}|, \qquad 1 \leq t < T.$$

Based on asymptotic arguments about the test statistic, Pettitt (1979) defines the following approximate p value of the test:

$$p \approx 2\exp\left(\frac{-6K_\tau^2}{T^3 + T^2}\right). \tag{1}$$

Thus, given a significance level α, if p value < α, the null hypothesis, H0, that the two distributions are equal is rejected.

## 2.1 EMPIRICAL ERROR RATE IN STATISTICAL HYPOTHESIS TESTING

Every hypothesis test is subject to two types of errors, namely: Type I error and Type II error. The Type I error is established when the null hypothesis H0 is rejected when it is true. The Type II error occurs when the null hypothesis H0 is not rejected when it is false (Montgomery and Runger, 2006). In hypothesis testing, the probability of the Type I error is controlled by the α, while the Type II error is usually not controlled. The probability of the Type I error is the size of the statistical test and, from the Type II error, the power of the test is evaluated in rejecting H0 given that it is false (Rizzo, 2007). Thus, the performance evaluation of a hypothesis test is usually given through the size and power of the test.

The size and power of tests can only be determined when the real distribution and parameters of the data are known. In real data, this information is unknown, but the chances of occurrence of errors decrease as the analyzed sample size becomes larger (Walpole et al., 2006). However, long series are not always possible. Therefore, it is preferable to use tests with a lower probability of error, that is, a greater detection power for a fixed test size. However, there are situations in which the test is distorted in size. The size distortion occurs through the incompatibility of the pre-assigned significance level of the test and the real probability of the Type I error. In a test whose size is lower than the predefined significance level, the Type II error probability is greater than it should be, therefore a reduction in the test's detection power occurs. On the other hand, if you are using a test whose size is greater than the pre-specified significance level, the opposite occurs, the probability of the Type II error decreases and the detection power of the test artificially increases.

Correcting the size distortion of a test ensures that the rejection rate of the null hypothesis, when it is true, actually corresponds to the significance level used. From the size correction of the test, a "real" detection power is obtained, that is, without increasing or decreasing the rejection rate of the test due to the influence of the size distortion. In this sense, we propose a bootstrap application for the Pettitt test, since its p value, given in (1), arises from asymptotic arguments and can be quite distorted in small and moderate sample sizes. By means of the proposed test, it is sought an improvement of the performance of the test, reducing errors, and increasing the detection power. In the following sections, the proposed test is presented and evaluated.

## 3 BOOTSTRAP PETTITT TEST

The proposed test is based on the bootstrap method (Davison and Hinkley, 1997; Efron and Tibshirani, 1994), by means of the resampling of a data series and in empirical calculations from these bootstrap resamples. The resampling of the data is made from the original sample, treating it as a pseudo-population. Having the empirical distribution of the test statistic under H0, the test calculates the bootstrap p value (Davison and Hinkley, 1997). This calculated p value (pboot) is based on the test statistic of the original sample compared to the statistics obtained with the resampling. The following algorithm defines the proposed test (P.T*):

1. Given the observed sample X = (x1,…, xT ), we compute the Pettitt test statistic (Kτ).

2. We generate B bootstrap resamples of the series, X*b = (x*b1, ..., x*bT ), with b = 1, …, B, being established with replacement and with sample size T.
3. For each bootstrap resample X*b the test statistic is calculated. Thus, a statistic (Kτ*b) for each resample is obtained.
4. The bootstrap p value is determined by:

$$p_{boot} = \frac{1 + \sum_{b=1}^{B}\{|K_\tau^{*b}| \geq |K_\tau|\}}{B + 1}.$$

5. We reject H0 if pboot < α, for a fixed significance level α. Usually α = 0.01, 0.05 or 0.10.

The objective of the proposed test is to test the hypothesis of no change in a data series. If the corrected test decides by the presence of the change point, the location of this point is given by the classical Pettitt test.

# 4 NUMERICAL EVALUATION

In this section the size and power of the Pettitt test (P.T) and of the proposed test (P.T*) are evaluated through Monte Carlo simulation. We considered R = 10 000 Monte Carlo replications and B = 1 000 bootstrap resamples. The simulation study was performed using the R language (R Core Team, 2017).

## 4.1 SIMULATION SCENARIOS

For the evaluation, we considered synthetic series of independent continuous random variables with positive values in order to represent processes such as precipitation, flow, relative humidity, among others, such as in of Xie et al. (2014) and Mallakpour and Villarini (2016). The mean value adopted to the stationary series was m =100, since it approximates the annual average monthly precipitation (in mm) of the climatological normals for regions such as Northeast and Southeast of Brazil (INMET, 1992).

These synthetic series were specified using three different probability distributions, gamma, Gumbel and normal. These distributions were chosen due to their importance in hydroclimatic time series studies. The gamma distribution has the advantage of presenting only positive values, which occurs in some hydrological series such as precipitation and flow, having good applicability in these cases (Aksoy, 2000; Yoo et al., 2005; Yue et al., 2001). The Gumbel distribution was chosen because it has applicability in simulations of extreme hydroclimatological events (Leese, 1973; Yue et al., 1999). Finally, we considered the normal distribution, which is often the required distribution of the data in the applicability of parametric tests. In Appendix A we describe the probability density functions of the distributions used for this simulation study. In addition to the different distributions used to generate the synthetic series, we also considered some different scenarios, with: (i) five different sample size (T), (ii) six different magnitudes of the shift (S), (iii) four different coefficients of variation (CV), and (iv) three different location of the change point (τ).

Regarding the sample size, we evaluated from short to long series, with T =10, 20, 30, 50, and 100. To characterize the change point, defined in time τ, we subdivided the series in two new sub-series with different means. The difference between the means was defined by the magnitude of the shift (S), evaluated in percentage terms of the mean. We insert the change in the monotonous mean with the following magnitude variations: $S = \pm 10\%$, $5\%$, $3\%$, $1\%$, $0.1\%$ and $0\%$. In the case of $S = 0\%$ there is no difference between the means of the sub-series, being useful to evaluate the size of the test. The generated sub-series are defined as:

$$\text{mean} = \begin{cases} \mu, & \text{when } t = 1, \dots, \tau, \\ \mu + (S \times \mu), & \text{when } t = \tau + 1, \dots, T. \end{cases}$$

In addition to the magnitude of the shift S in the series, we evaluated the influence of the data variance on the power of the test. For this purpose, we considered the following coefficients of variation: $CV = 5\%$, $10\%$, $20\%$ and $30\%$. As for the position of the change point $t$ in the series, we consider $\tau$ located at 10% (a sub-series with 10% of the sample data and the other one with 90%), 50% and 70% of the sample size.

For the different scenarios, we investigated the number of times (Rdet) the Pettitt tests, with and without bootstrap application, identified the change point. Thus, we set the test rejection rate as follows:

$$\text{Rejection rate} = \frac{R_{\text{det}}}{R},$$

where R is the total number of simulated replications (R = 10 000), evaluated under significance level α. The significance levels considered were $\alpha = 0.01$, 0.05 and 0.10.

## 4.2 SIMULATION RESULTS

Table 1 shows the results of the numerical analysis of size distortion for the gamma distribution, where P.T refers to the Pettitt test and P.T* to the proposed test. In the size evaluation, the results are independent of the change point location t, since they are generated under H0, that is, without a change point ($\tau = 0$). In terms of size, regardless of the distribution assessed, the classical Pettitt test showed large distortions, especially for small samples. The results obtained for the uncorrected test are close to those presented by Xie et al. (2014) and Mallakpour and Villarini (2016). For the proposed test, small distortions were obtained in all scenarios, keeping the values very close to the nominal values of α. For brevity and similarity of results, the results of the size evaluation for the other distributions are found in Appendix B.

Table 1 – Evaluation of the Pettitt test size (P.T) and of the proposed test size (P.T*) for gamma distribution, where CV is the coefficients of variation, α is the significance level and T is the sample length.

| CV | 5% | | 10% | | 20% | | 30% | |
|---|---|---|---|---|---|---|---|---|

*α = 0.01*

| | P.T | P.T* | P.T | P.T* | P.T | P.T* | P.T | P.T* |
|---|---|---|---|---|---|---|---|---|
| *T*=10 | 0.0000 | 0.0084 | 0.0000 | 0.0077 | 0.0000 | 0.0088 | 0.0000 | 0.0087 |
| *T*=20 | 0.0014 | 0.0100 | 0.0008 | 0.0094 | 0.0010 | 0.0110 | 0.0012 | 0.0072 |
| *T*=30 | 0.0032 | 0.0099 | 0.0031 | 0.0093 | 0.0027 | 0.0101 | 0.0030 | 0.0093 |
| *T*=50 | 0.0039 | 0.0101 | 0.0028 | 0.0073 | 0.0049 | 0.0099 | 0.0043 | 0.0099 |
| *T*=100 | 0.0060 | 0.0092 | 0.0081 | 0.0125 | 0.0076 | 0.0109 | 0.0057 | 0.0103 |

*α = 0.05*

| | P.T | P.T* | P.T | P.T* | P.T | P.T* | P.T | P.T* |
|---|---|---|---|---|---|---|---|---|
| *T*=10 | 0.0000 | 0.0455 | 0.0000 | 0.0446 | 0.0000 | 0.0437 | 0.0000 | 0.0451 |
| *T*=20 | 0.0180 | 0.0499 | 0.0178 | 0.0471 | 0.0190 | 0.0494 | 0.0180 | 0.0488 |
| *T*=30 | 0.0247 | 0.0524 | 0.0243 | 0.0485 | 0.0251 | 0.0504 | 0.0228 | 0.0496 |
| *T*=50 | 0.0296 | 0.0498 | 0.0288 | 0.0504 | 0.0306 | 0.0529 | 0.0292 | 0.0496 |
| *T*=100 | 0.0355 | 0.0497 | 0.0390 | 0.0534 | 0.0361 | 0.0507 | 0.0333 | 0.0471 |

*α = 0.10*

| | P.T | P.T* | P.T | P.T* | P.T | P.T* | P.T | P.T* |
|---|---|---|---|---|---|---|---|---|
| *T*=10 | 0.0253 | 0.0953 | 0.0205 | 0.0941 | 0.0267 | 0.0994 | 0.0243 | 0.0921 |
| *T*=20 | 0.0408 | 0.0923 | 0.0432 | 0.0958 | 0.0452 | 0.0970 | 0.0442 | 0.0973 |
| *T*=30 | 0.0527 | 0.1022 | 0.0523 | 0.1000 | 0.0526 | 0.0987 | 0.0509 | 0.0963 |
| *T*=50 | 0.0611 | 0.1004 | 0.0636 | 0.1007 | 0.0633 | 0.1026 | 0.0611 | 0.0964 |
| *T*=100 | 0.0737 | 0.0990 | 0.0790 | 0.1035 | 0.0757 | 0.1014 | 0.0811 | 0.1048 |

As shown in Table 1, the size of the Pettitt test is zero when T = 10, α = 0.01 and 0.05. This reduction in the probability of the Type I error may seem advantageous; however, as commented in Section 2.1, this result causes the probability of the Type II error to be higher, obtaining a test with reduced change detection power. From the results of Table 1, the size distortion of the Pettitt test in small sample sizes is evident when compared to the results of the proposed test. We realize that, in T = 30 and α = 0.10, when applying a test at a significance level of 10%, the classical Pettitt test size is 23%, while the corrected test size is exactly 10%. These distortions for the classical Pettitt test decrease the greater the number of data points in the series, while the corrected test remains with size always close to the nominal value of α, regardless of the sample size. Therefore, in terms of size, when compared to the classical Pettitt test, the proposed test is noticeably more suitable for evaluation of the change point.

With the size correction of the proposed test, we obtained a test power subject to a probability of the Type I error closest to the defined nominal value of α. Following are the results from the detection power evaluation of the considered tests. Figures 1 and 2 illustrate the comparison of the power of the Pettitt test and of the proposed test in the identification of the change point for the three distributions evaluated, under the different levels of changes. This power assessment is presented for a sample size of T = 50, a coefficient of variation CV = 5%, and significance level α=0.05.

Figure 1: Power of the Pettitt test in the evaluated distributions; T=50 and CV=5%. (a) Change point in 10%; (b) Change point in 50%; (c) Change point in 70%.

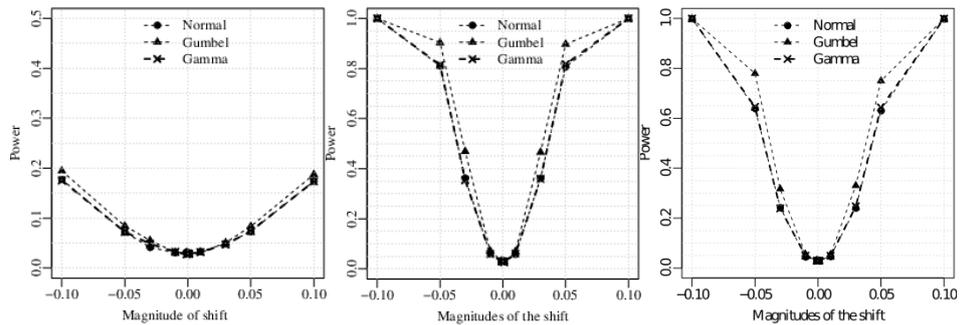

Figure 2: Power of the proposed test in the evaluated distributions; T=50 and CV=5%. (a) Change point in 10% ; (b) Change point in 50%; (c) Change point in 70%.

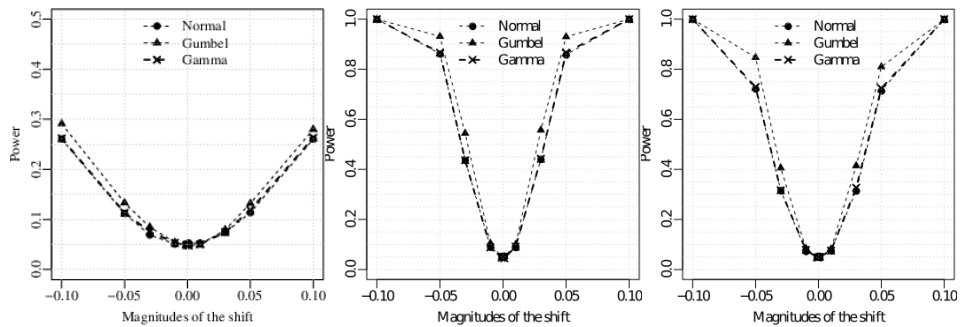

According to the results of the numerical simulations, the power of the tests varies little among the distribution of the evaluated series. This result was already expected since the Pettitt test is based on rankings, that is, redistribution of the data according to its position within the series in terms of degree classification, being nonparametric. Therefore, the next results will be based only on the gamma distribution.

Figure 3 shows the power of both tests, under different simulated significance levels. We note the power of the proposed test is greater than or equal to the power of the classical test in all scenarios. We can emphasize that the larger the sample size and the higher the significance level, the better is the detection power of the change point. This happens because as α increase, the allowed rejection rate gets larger. For the discussion of power that follows, we consider only α = 0.05.

Figure 3: Power of the tests for the different significance levels (α); gamma distribution, CV=5%, location point 50% and magnitude of the shift S=5% of the mean. (a) α = 0.01; (b) α = 0.05 ; (c) α = 0.1.

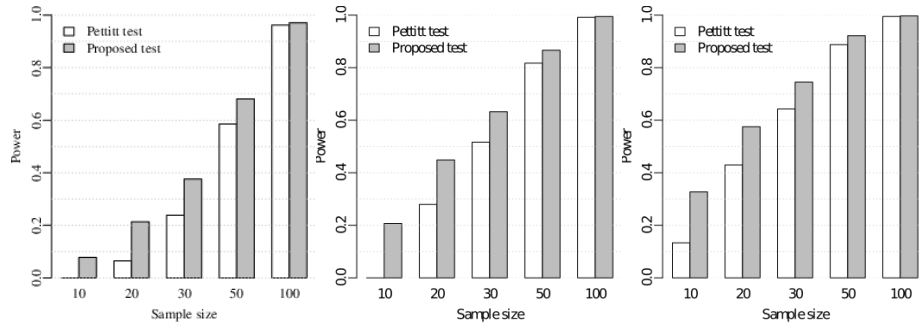

A comparison of the power of the tests under different change point position and the sample size is shown in Figure 4. These results consider S = 5% and CV = 5%. Power evaluation results for others magnitudes of the shift are presented in Appendix C. Again, the power of the proposed test is greater than or equal to the power of the classical Pettitt test. Thus, it is possible to verify that even with the correction of the test size, the power is not impaired. In addition, for small sample sizes (T = 10, 20, or 30), regardless of the change point location, the proposed test performs significantly better in identifying the change point. Note that when T = 10, the Pettitt test has power equals to zero; so, in the 10 thousand synthetic series, the classical test did not identify any significant change in the series.

Figure 4: Detection power of the tests for the different change point locations; gamma distribution, CV=5%, α=0.05 and magnitude of the sift S=5% of the mean. (a) Change point in 10%; (b) Change point in 50%; (c) Change point in 70%.

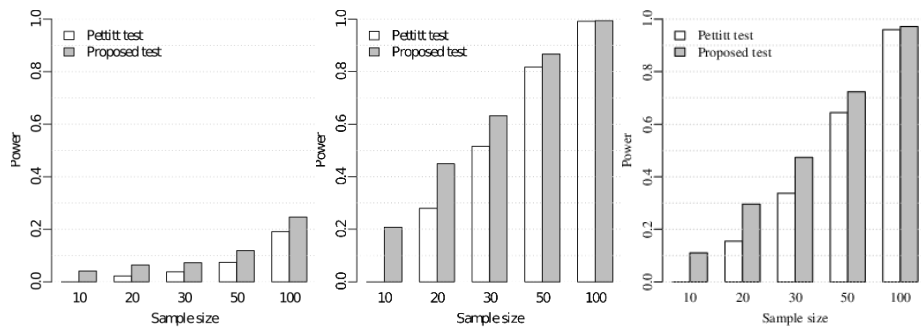

In Figure 4 it is also possible to identify that the closer to the center of the series the point is, the easier it is for the tests to detect the change point, reducing this power in the extremities. This characteristic had already been identified in the studies carried out by Xie et al. (2014) and Mallakpour and Villarini (2016) for the Pettitt test. However, the proposed test is always more powerful.

In the same way as for the change magnitude of S = 5% (Figure 4), the classical Pettitt test did not present satisfactory results for the other simulated magnitudes, especially for small sample sizes, regardless of the change point location (Appendix C, Figures 8 to 11). For example, for a sample size of T = 10 and a magnitude of the shift of S = 10% , the proposed test reached rates of 60% of detection, with the location point at the center of the distribution and coefficient of variation equals to 5%, whereas the classical test did not detect any change (Appendix C, Figure 8(b)).

Figure 5 indicates the power comparison of the tests in identifying the change point for different coefficients of variation. The greater the variance of the data, the more difficult it is to identify the change point for both tests. In addition, the proposed test has greater detection power than the classical test.

Figure 5: Detection power of the tests for the different coefficients of variation; gamma distribution, T=50, α=0.05 and magnitude of the sift S=5% of the mean. (a) Change point in 10%; (b) Change point in 50%; (c) Change point in 70%.

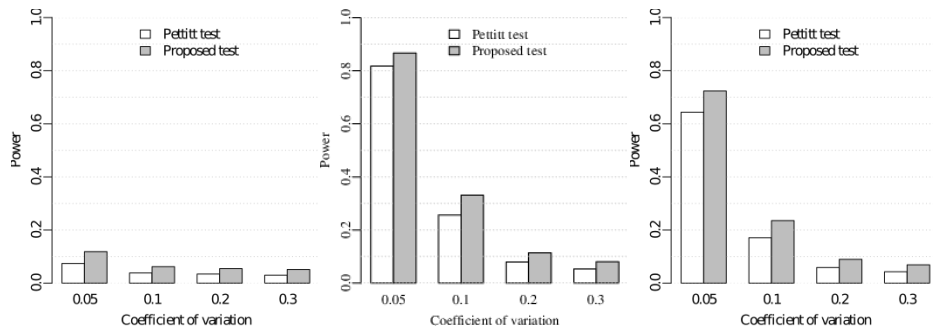

With these results, it is possible to identify relations between detection power, sample size, the magnitude of the shift, coefficient of variation, and the change point location. Both the classical Pettitt test and the proposed test easily identify the change point the larger the sample size, the larger the magnitude of the shift, the smaller the variability of the data, and the closer to the center of the series this point is located. In general, the proposed test was able to correct the size distortions while showing greater detection power than the classical test.

## 5 CASE STUDY

Anthropogenic actions, such as deforestation and replacement of natural pastures by agriculture, have become a common practice in several basins. In addition, the global warming, which has been projected by climate models, includes increases in mean temperature in most land and ocean regions, in hot extremes in most inhabited regions, in heavy precipitation in several regions, and in the probability of drought and precipitation deficits in some regions (IPCC, 2018). These natural and anthropogenic changes are directly linked on the streamflows. In this case study, we analyzed the annual monthly averages flow of the Paraná River of the La Plata basin, where several studies have shown a change point in the 70s (Boulanger et al., 2005; García and Vargas, 1998; García et al., 2002; Puig et al., 2016; Tucci and Clarke, 1998).

The La Plata basin, 2nd largest river basin in South America, is important in social and economic issues, mainly because it is densely populated, with one of the most important agricultural and hydroelectricity production regions in the world (Bettolli and Penalba, 2018; Boulanger et al., 2005; Cavalcanti et al., 2015; Doyle and Barros, 2011; Montroull et al., 2018; Tucci and Clarke, 1998). In this basin is located the Itaipu Hydroelectric plant with 14,000 mW installed power (ITAIPU, 2018). Water resources of the La Plata basin are therefore directly related to the sustainable development of the region, where a large fraction of the economic activities depends on the water availability (Cavalcanti et al., 2015; Doyle and Barros, 2011).

La Plata basin is a good case where identifying a change point is important to understand its causes, and support decision making in relation to water management.

We considered the naturalized flow data of the Itaipu Hydroelectric plant, in Brazil, from 1931 to 2015. The naturalized flow is the natural flow in the river without some anthropogenic action in the basin, such as reservoir regularization, flow transposition, or water abstraction (ONS, 2019). This series was obtained from the database of the National Electric System Operator (ONS, 2017), which is the agency responsible for coordinating and controlling the operation of electric power generation and transmission in Brazil. Figure 6 shows the map of the considered fluviometric station.

Figure 6: Map of the studied fluviometric station.

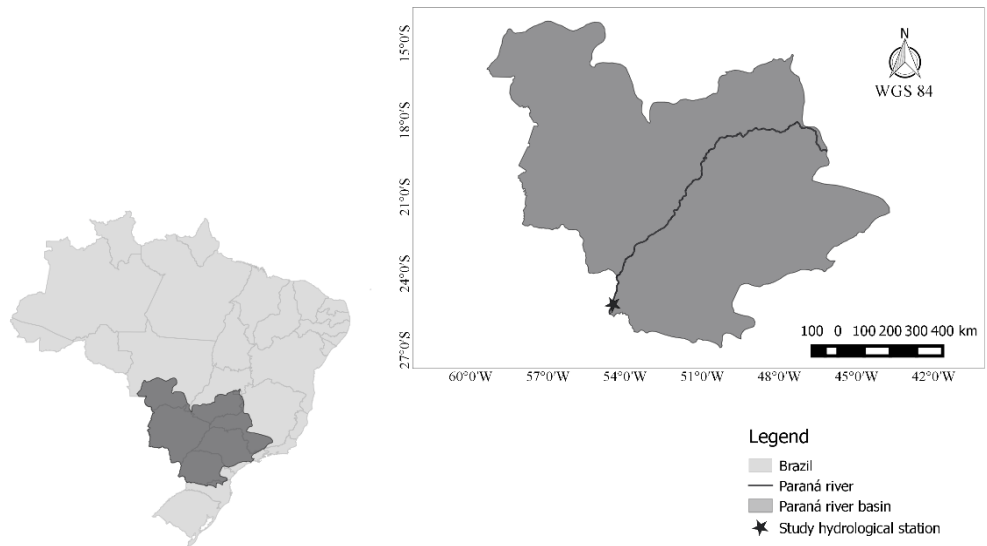

The mean of this series of flow data is 10 319.99 m³/s with standard deviation of 2 522.07 m³/s. The sample autocorrelation function (ACF) (Figure 7(a)) and the sample partial autocorrelation function (PACF) (Figure 7(b)) indicate nonnull autocorrelations in the time series. Nevertheless, the Pettitt test supposes independence in the data set to its application. In this way, we used a prewhitening method adapted from Serinaldi and Kilsby (2016) in the actual flow time series. This methodology consists of the following steps:

1. The Pettitt test is applied to the original data. If there is no evidence to reject H0, then there is "no change" and the analysis stopped.
2. If there is evidence to reject H0, then the times series is split in two sub-series, $t = 1, \ldots, \tau$ and $t = \tau + 1, \ldots, T$. The difference of the means is computed by $\Delta = \mu_b - \mu_a$, and used to remove the step $x_t = y_t - \Delta . \mathbf{1}_{\{t > \tau\}}$.
3. The autocorrelation is removed by $\varepsilon_t = x_t - \hat{\rho}^* x_{t-1}$, where

$$\hat{\rho}^* = \left(\hat{\rho} + \frac{1}{T}\right)\left(\frac{T}{T-3}\right),$$

with

$$\hat{\rho} = \frac{\dfrac{1}{(T-1)}\sum_{t=1}^{(T-1)}(X_t - \hat{\mu})(X_{t+1} - \hat{\mu})}{\dfrac{1}{T}\sum_{t=1}^{T}(X_t - \hat{\mu})^2},$$

and

$$\hat{\mu} = \frac{1}{T} \sum\nolimits_{t=1}^{T} X_t.$$

4.  Finally, the step change is combined with the uncorrelated series

$$\hat{\Delta}. 1_{\{t>\tau\}} + \frac{\varepsilon_t}{1 - \hat{\rho}^{*}},$$

and the classical and proposed Pettitt tests are applied to this prewhitened series.

Figure 7: Observed ACF and PACF of flow data (a)ACF; (b) PACF.

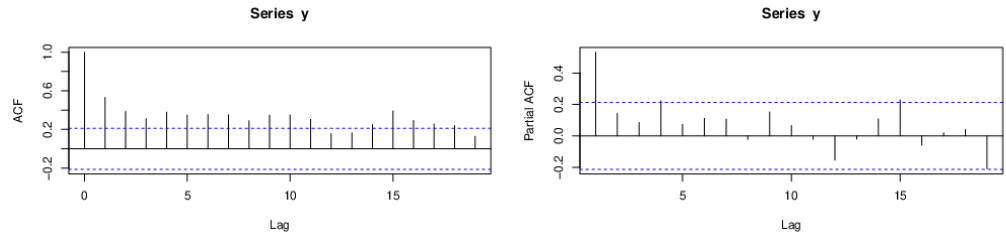

By applying the Pettitt test and the proposed test in the complete series of flow data (T = 85), we obtained the identification of the change point for both tests, with p values lower than 0.01. The change point was identified as being in 1971, as shown in Figure 8. This result corroborates with the abovementioned papers, and the change in natural flow may has been caused by deforestation processes, intensive agriculture introduction, urban development (Tucci and Clarke, 1998), or climate changes. As expected, we note a good performance of the classical Pettitt test in this application because (i) the sample size is large (T = 85), (ii) the magnitude of the shift is big (S = 38.96%), and (iii) the change point is almost in the midpoint of the series ($\tau = 48\%$).

Figure 8: Monthly average naturalized flow by year (continuous line) and average naturalized flow by long term for the period before and after 1971 (dashed line).

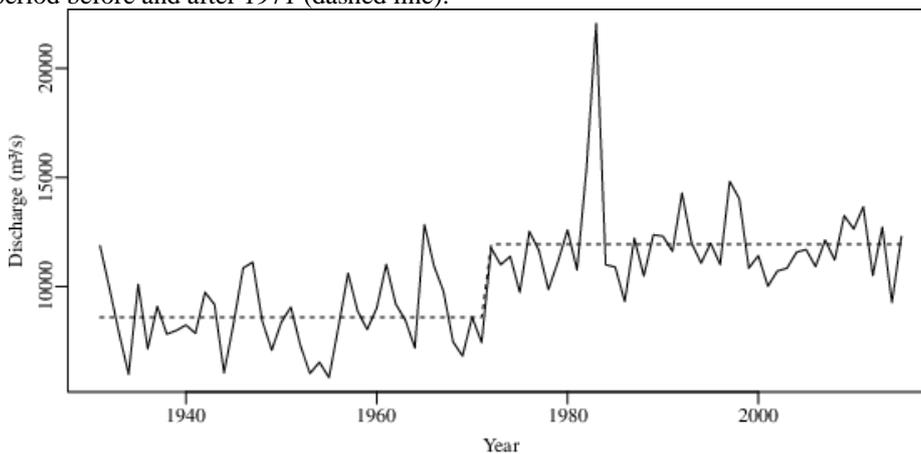

However, a small data series is a reality in many monitoring stations, where the change point is not always located in the center of the series. In order to mimic other real situations, we considered five different scenarios making cuts in the original data set. These scenarios are named by C1, C2, C3, C4, and C5, and their periods are described in Table 2. The scenario C1 is the only one that does not include the level change previously detected in the complete data. The C2 presents the largest sample size among the five scenarios. C5 has the smallest sample sizes (T = 32). Scenarios C4 and C5 are similar in relation to CV, $\tau$, and S, but they present different sample sizes. The CV of the data in all scenarios ranges from 19.3% to 28.7 %. The results of tests application in each scenario are also presented in Table 2. For both tests, the classical Pettitt test and the proposed test, we considered significance levels equal to 5% (α = 0.05). The classical Pettitt test and the improved test did not reject the null hypothesis for C1 in step 1 of the Serinaldi and Kilsby (2016) methodology, then the analysis is stopped.

Table 2 - Application of the classical Pettitt test (P.T) and the proposed test (P.T*) in different samples from the Itaipu annual monthly flow data. The CV is the coefficient of variation of the sample, the change point column presents the position of the change point τ (%) in the series and the year of this change, and S is the magnitude of the shift.

| | Periods | $T$ (sample size) | $CV$ (%) | P.T | | Proposed P.T* | | Change point ($\tau$) | | $S$ (%) |
|---|---|---|---|---|---|---|---|---|---|---|
| | | | | $p$ value | Detection | $p$ value | Detection | (%) | Year | |
| C1 | 1931-1969 | 39 | 19.3 | 0.68831[1] | NO | 0.53247[1] | NO | - | - | - |
| C2 | 1931-1990[2] | 60 | 27.1 | 0.00101 | **YES** | 0.00099 | **YES** | 68 | 1971 | 40.1 |
| C3 | 1945-1986[2] | 42 | 28.7 | 0.07147 | NO | 0.04196 | **YES** | 64 | 1971 | 39.8 |
| C4 | 1961-2015[2] | 55 | 20.5 | 0.01296 | **YES** | 0.00499 | **YES** | 30 | 1977 | 21.8 |
| C5 | 1965-1996[2] | 32 | 23.5 | 0.09032 | NO | 0.04795 | **YES** | 39 | 1977 | 21.8 |

[1] p value of P.T and P.T* to the original data.
[2] with change point detected in the complete data series.

The results presented in Table 2 show that the classical Pettitt test did not reject the null hypothesis for C3 (from 1945 to 1986) and C5 (from 1965 to 1968), i.e., it did not identify significant change in the mean of the series, unlike the test that uses the bootstrap method. In both scenarios, the time series includes the change point, the year 1971, in the extremity of the series. The proposed test was able to detect this change point, with p value lower than 5%. These cases highlight that the detection performance of the proposed test is greater than the classical Pettitt test for short series, especially when the change point is far from the midpoint. Both tests reject the null hypothesis for C2 and C4. These scenarios are related to the large sample size, so it was expected that classical Pettitt test is able to identify the change. This result is in accordance with the one previously presented by Monte Carlo simulation.

It should be noted that in real data applications it is not always possible to choose the best range and sample size of the data. Then, if an inadequate test is applied, the results could lead to the wrong conclusions about the presence or not of the change points.

## 6 CONCLUSIONS

In this study, we proposed an application of the Pettitt test via the bootstrap method. We examined the performance of both tests, the classical and proposed Pettitt tests, in identifying change point in synthetic hydroclimatological data series using different scenarios through the Monte Carlo simulation. We defined scenarios in order to incorporate a wide spectrum of possibilities, easily found in real data and with the view to contemplate short series, already identified as inadequate for the application of the classical Pettitt test.

The simulation results identified the size distortion of the Pettitt test, which was corrected or minimized through the bootstrap method application. By correcting the test size we obtained a more accurate and reliable detection, with Type I error probability closer to the significance level. Even with the correction of size distortions, the proposed test is also more powerful than the classical Pettitt test for detection of change points.

As a case study, we applied this to the flow data from the Itaipu Hydroelectric plant. The result of this application corroborates with the simulation results. The bootstrap Pettitt test can be useful in real situations because it is a more reliable test than the classical one, mainly in small samples. When the hydroclimatological data have an important role for society it is important to quickly detect an abrupt change in order to support decision making for hydroclimatological resources management.

## AN R IMPLEMENTATION

An implementation in R language of the bootstrap Pettitt test is available at https://github.com/fabiobayer/bootpettitt

## ACKNOWLEDGEMENTS
We gratefully acknowledge partial financial support from Conselho Nacional de Desenvolvimento Científico e Tecnologico (CNPq) and Fundacão de Amparo a Pesquisa do Estado do Rio Grande do Sul (FAPERGS), Brazil.

## APPENDIX A

The probability distributions used in the simulation study described in Section 4 are described below.

### A.1. Gamma distribution

Let X be a random variable with a gamma distribution, the probability density function of X is given by (Montgomery and Runger, 2006):

$$f(x, \lambda, \theta) = \frac{\lambda^{\theta} x^{\theta-1} e^{-\lambda x}}{\Gamma(\theta)}, \qquad x > 0,$$

where $\lambda > 0$ and $\theta > 0$ are the parameters, $E(X) = \frac{\theta}{\lambda}$ and $\text{Var}(X) = \frac{\theta}{\lambda^2}$.

### A.2. Gumbel distribution

The probability density function of the Gumbel distribution is given by (Chow, 1964):

$$f(x; \lambda, \theta) = \frac{1}{\lambda} \exp\left[-\frac{1}{\lambda}(x - \theta) - \exp\left[-\frac{1}{\lambda}(x - \theta)\right]\right], \qquad -\infty < x < \infty,$$

where $\lambda > 0$ and $-\infty < x < \infty$ are the parameters, $E(X) = \theta + \frac{\gamma}{\lambda}$ and $\text{Var}(X) = \frac{\pi^2}{6\lambda^2}$, being $\gamma$ the Euler constant given by approximately 0.577.

### A.3. Normal distribution

The normal distribution is one of the most usual distribution in statistical modeling, assuming a symmetric bell shape (Walpole et al., 2006). It has as a characteristic a data distribution with support from $-\infty < x < \infty$, which is the fact that leads to the exclusion of several hydrological variables such as flow, precipitation and humidity, because there is no possibility of negative observations. The probability density function of this distribution is given by (Montgomery and Runger, 2006):

$$f(x; \mu, \sigma^2) = \frac{1}{\sigma\sqrt{2\pi}} \exp\left[\frac{-(x - \mu)^2}{2\sigma^2}\right], \qquad -\infty < x < \infty,$$

where $-\infty < x < \infty$ is the mean and $\sigma^2 > 0$ is the variance.

## APPENDIX B

Tables 3 and 4 present the results obtained for Gumbel and normal distribution, respectively (cf. Table 1 for gamma distribution).

Table 3 - Evaluation of the Pettitt test size (P.T) and of the proposed test size (P.T*) for Gumbel distribution, where CV is the coefficients of variation, α is the significance level and T is the sample length.

| | Gumbel distribution | | | | | | | |
|---|---|---|---|---|---|---|---|---|
| *CV* | 5% | | 10% | | 20% | | 30% | |
| | | | | α = 0.01 | | | | |
| | P.T | P.T* | P.T | P.T* | P.T | P.T* | P.T | P.T* |
| *T*=10 | 0.0000 | 0.0080 | 0.0000 | 0.0076 | 0.0000 | 0.0090 | 0.0000 | 0.0084 |
| *T*=20 | 0.0015 | 0.0093 | 0.0017 | 0.0094 | 0.0011 | 0.0097 | 0.0018 | 0.0105 |
| *T*=30 | 0.0031 | 0.0112 | 0.0033 | 0.0100 | 0.0027 | 0.0095 | 0.0029 | 0.0090 |
| *T*=50 | 0.0035 | 0.0091 | 0.0048 | 0.0104 | 0.0046 | 0.0113 | 0.0044 | 0.0093 |
| *T*=100 | 0.0059 | 0.0099 | 0.0053 | 0.0101 | 0.0057 | 0.0098 | 0.0046 | 0.0080 |
| | | | | α = 0.05 | | | | |
| | P.T | P.T* | P.T | P.T* | P.T | P.T* | P.T | P.T* |
| *T*=10 | 0.0000 | 0.0441 | 0.0000 | 0.0417 | 0.0000 | 0.0436 | 0.0000 | 0.0412 |
| *T*=20 | 0.0168 | 0.0497 | 0.0180 | 0.0505 | 0.0186 | 0.0477 | 0.0202 | 0.0534 |
| *T*=30 | 0.0230 | 0.0489 | 0.0250 | 0.0506 | 0.0229 | 0.0483 | 0.0205 | 0.0470 |
| *T*=50 | 0.0290 | 0.0482 | 0.0316 | 0.0531 | 0.0295 | 0.0498 | 0.0290 | 0.0518 |
| *T*=100 | 0.0353 | 0.0486 | 0.0332 | 0.0485 | 0.0368 | 0.0519 | 0.0382 | 0.0527 |
| | | | | α = 0.10 | | | | |
| | P.T | P.T* | P.T | P.T* | P.T | P.T* | P.T | P.T* |
| *T*=10 | 0.0240 | 0.0937 | 0.0225 | 0.0933 | 0.0233 | 0.0959 | 0.0226 | 0.0895 |
| *T*=20 | 0.0463 | 0.1019 | 0.0471 | 0.0969 | 0.0439 | 0.1009 | 0.0492 | 0.1029 |
| *T*=30 | 0.0508 | 0.0962 | 0.0523 | 0.1021 | 0.0533 | 0.1006 | 0.0524 | 0.0981 |
| *T*=50 | 0.0578 | 0.0940 | 0.0683 | 0.1049 | 0.0647 | 0.1009 | 0.0647 | 0.1014 |
| *T*=100 | 0.0723 | 0.0994 | 0.0767 | 0.1038 | 0.0745 | 0.0994 | 0.0745 | 0.0988 |

Table 4 - Evaluation of the Pettitt test size (P.T) and the proposed test (P.T*) for normal distribution, where CV is the coefficients of variation, α is the significance level and T is the sample length.

| Normal distribution | | | | | | | |
|---|---|---|---|---|---|---|---|
| *CV* | 5% | | 10% | | 20% | | 30% |

| α = 0.01 | | | | | | | |
|---|---|---|---|---|---|---|---|
| | P.T | P.T* | P.T | P.T* | P.T | P.T* | P.T | P.T* |
| *T*=10 | 0.0000 | 0.0100 | 0.0000 | 0.0075 | 0.0000 | 0.0080 | 0.0000 | 0.0078 |
| *T*=20 | 0.0010 | 0.0092 | 0.0015 | 0.0093 | 0.0015 | 0.0093 | 0.0012 | 0.0094 |
| *T*=30 | 0.0015 | 0.0077 | 0.0027 | 0.0092 | 0.0029 | 0.0105 | 0.0035 | 0.0102 |
| *T*=50 | 0.0040 | 0.0112 | 0.0044 | 0.0106 | 0.0037 | 0.0099 | 0.0051 | 0.0113 |
| *T*=100 | 0.0053 | 0.0096 | 0.0050 | 0.0073 | 0.0061 | 0.0098 | 0.0065 | 0.0113 |

| α = 0.05 | | | | | | | |
|---|---|---|---|---|---|---|---|
| | P.T | P.T* | P.T | P.T* | P.T | P.T* | P.T | P.T* |
| *T*=10 | 0.0000 | 0.0434 | 0.0000 | 0.0445 | 0.0000 | 0.0447 | 0.0000 | 0.0473 |
| *T*=20 | 0.0168 | 0.0460 | 0.0176 | 0.0491 | 0.0193 | 0.0518 | 0.0167 | 0.0454 |
| *T*=30 | 0.0237 | 0.0499 | 0.0211 | 0.0455 | 0.0212 | 0.0470 | 0.0236 | 0.0513 |
| *T*=50 | 0.0299 | 0.0508 | 0.0321 | 0.0503 | 0.0283 | 0.0471 | 0.0290 | 0.0491 |
| *T*=100 | 0.0329 | 0.0466 | 0.0336 | 0.0496 | 0.0343 | 0.0471 | 0.0393 | 0.0521 |

| α = 0.10 | | | | | | | |
|---|---|---|---|---|---|---|---|
| | P.T | P.T* | P.T | P.T* | P.T | P.T* | P.T | P.T* |
| *T*=10 | 0.0256 | 0.0990 | 0.0264 | 0.1016 | 0.0246 | 0.0974 | 0.0235 | 0.0933 |
| *T*=20 | 0.0464 | 0.0992 | 0.0438 | 0.0975 | 0.0453 | 0.1016 | 0.0469 | 0.1004 |
| *T*=30 | 0.0497 | 0.0949 | 0.0483 | 0.0952 | 0.0503 | 0.0918 | 0.0597 | 0.1037 |
| *T*=50 | 0.0630 | 0.1028 | 0.0634 | 0.0959 | 0.0572 | 0.0950 | 0.0638 | 0.0966 |
| *T*=100 | 0.0744 | 0.1008 | 0.0775 | 0.1021 | 0.0766 | 0.1027 | 0.0731 | 0.1003 |

# APPENDIX C

The results of comparing the power of the tests to the different magnitudes of the shift evaluated in this study are presented for T = 10 (Figure 9), T = 20 (Figure 10), T = 30 (Figure 11), T = 50 (Figure 12) and T = 100 (Figure 13). A significance level of α = 0.05 was established for all scenarios, with coefficient of variation equals to 5% and gamma distribution.

Figure 9 - Power of the tests for the different magnitudes of the shift evaluated; T = 10. (a) Change point in 10%; (b) Change point in 50%; (c) Change point in 70%.

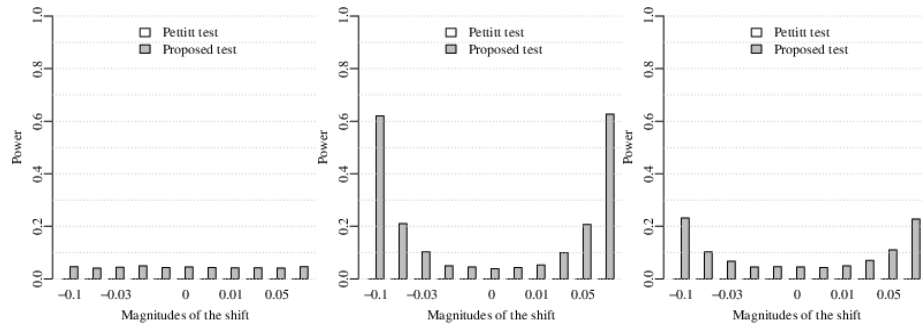

Figure 10 - Power of the tests for the different magnitudes of the shift evaluated; T = 20. (a) Change point in 10%; (b) Change point in 50; (c) Change point in 70%.

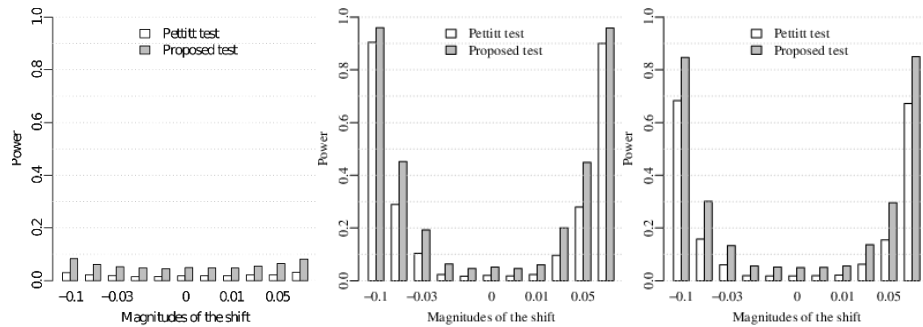

Figure 11 - Power of the tests for the different magnitudes of the shift evaluated; T = 30. (a) Change point in 10%; (b) Change point in 50%; (c) Change point in 70%.

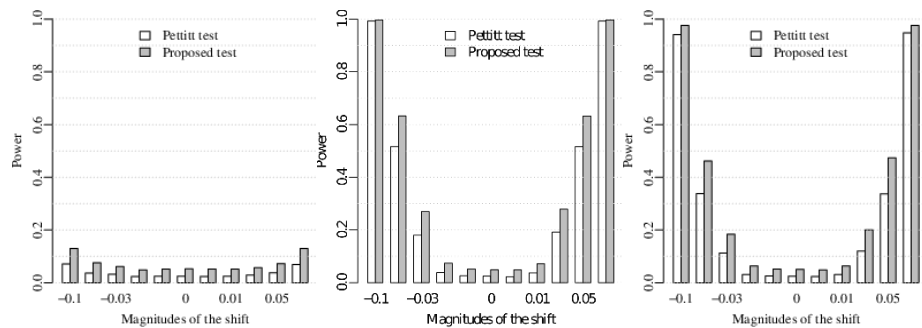

Figure 12 - Power of the tests for the different magnitudes of the shift evaluated; T = 50. (a) Change point in 10%; (b) Change point in 50%; (c) Change point in 70%.

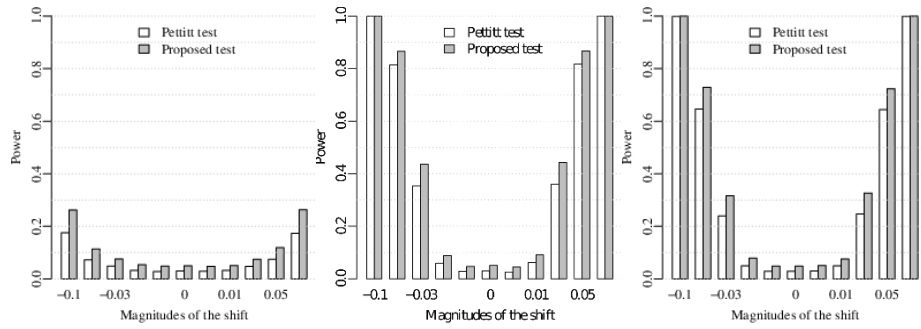

Figure 13 - Power of the tests for the different magnitudes of the shift evaluated; T = 100. (a) Change point in 10%; (b) Change point in 50%; (c) Change point in 70%.

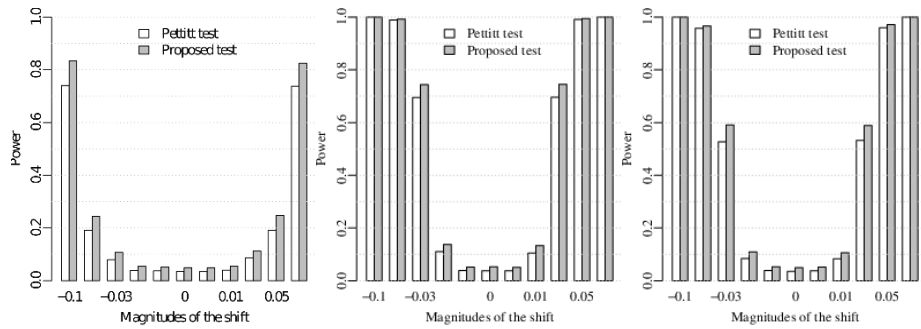